\documentstyle[multicol,eqsecnum,aps,psfig]{revtex}
\renewcommand{\narrowtext}{\begin{multicols}{2}
\global\columnwidth20.5pc\noindent}
\renewcommand{\widetext}{\end{multicols}
\global\columnwidth42.5pc}
\begin{document}
\draft
\preprint{1 May 2000}
\title{Multi-plateau magnetization curves of one-dimensional
       Heisenberg ferrimagnets}
\author{Shoji Yamamoto}
\address
{Department of Physics, Okayama University,
 Tsushima, Okayama 700-8530, Japan}
\author{T$\hat{\mbox o}$ru Sakai}
\address
{Faculty of Science, Himeji Institute of Technology,
 Ako, Hyogo 678-1297, Japan}
\date{Received 27 March 2000}
\maketitle
\begin{abstract}
Ground-state magnetization curves of ferrimagnetic Heisenberg chains
of alternating spins $S$ and $s$ are numerically investigated.
Calculating several cases of $(S,s)$, we conclude that the
spin-$(S,s)$ chain generally exhibits $2s$ magnetization plateaux
even at the most symmetric point.
In the double- or more-plateau structure, the initial plateau is
generated on a classical basis, whereas the higher ones are based on
a quantum mechanism.
\end{abstract}
\narrowtext

\section{Introduction}

   Ground-state magnetization curves of low-dimensional quantum spin
systems have been attracting much recent interest due to their
nontrivial appearance contrasting with classical behaviors.
A few years ago there appeared an epochal argument \cite{Oshi84} in
the field.
Generalizing the Lieb-Schultz-Mattis theorem \cite{Lieb07,Affl86},
Oshikawa, Yamanaka, and Affleck proposed that magnetization plateaux
of quantum spin chains should be quantized as
\begin{equation}
   S_{\rm unit}-m=\mbox{integer}\,,
   \label{E:OYA}
\end{equation}
where $S_{\rm unit}$ is the sum of spins over all sites in the unit
period and $m$ is the magnetization $M$ divided by the number of
the unit cells.
Their argument caused renewed interest \cite{Okam01} in the
pioneering calculations \cite{Zasp97,Okam39,Hida59} of a
bond-trimerized spin-$\frac{1}{2}$ chain and further stimulated
extensive investigations into quantum magnetization process.
Quantized magnetization plateaux were reasonably detected for
spin-$\frac{1}{2}$ \cite{Tots54}, spin-$1$
\cite{Tone17,Tots03,Naka26}, and spin-$\frac{3}{2}$
\cite{Saka01,Kita} chains with modulated and/or anisotropic
interactions.
Totsuka \cite{Tots05}, Cabra and Grynberg \cite{Cabr19}, and Honecker
\cite{Hone90} developed calculations of general polymerized spin
chains.
Numerous authors have been making further theoretical explorations
into extended systems including spin ladders
\cite{Cabr26,Tand96,Cabr68,Saka48} and layered magnets
\cite{Kole81,Hone97,Hone}.
Experimental observations \cite{Naru09,Shir48} have also followed.

   Mixed-spin chains, which have vigorously been studied in recent
years \cite
{Dril13,Alca67,Tian53,Pati94,Ono76,Nigg31,Ivan24,Kura62,Kole36,Wu57,Yama42},
also stimulate us in this context.
Theoretical investigations into them are all the more interesting and
important considering an accumulated chemical knowledge on
ferrimagnetic materials.
Kahn {\it et al.} \cite{Kahn89} succeeded in synthesizing a series of
bimetallic chain compounds such as
MM$'$(pba)(H$_2$O)$_3$$\cdot$$2$H$_2$O
(pba $=$ $1,3$-propylenebis(oxamato)
 $=$ C$_7$H$_6$N$_2$O$_6$) and
MM$'$(pbaOH)(H$_2$O)$_3$
(pbaOH $=$ $2$-hydroxy-$1,3$-propylenebis(oxamato)
 $=$ C$_7$H$_6$N$_2$O$_7$),
where the alternating magnetic ions M and M$'$ are flexible
variables and therefore the low-dimensional ferrimagnetic behavior
could systematically be observed.
Caneschi {\it et al.} \cite{Cane76}
synthesized another series of mixed-spin chain compounds of general
formula M(hfac)$_2$NITR, where metal ion complexes M(hfac)$_2$ with
hfac $=$ hexafluoroacetylacetonate are bridged by nitronyl nitroxide
radicals NITR.
There also exist purely organic molecule-based ferrimagnets
\cite{Shio42}, where sufficiently small magnetic anisotropy, whether
of exchange-coupling type or of single-ion type, is advantageous for
observations of essential quantum mixed-spin phenomena.

Magnetization curves of Heisenberg ferrimagnetic chains were revealed
by Kuramoto \cite{Kura62}.
His argument covered an effect of next-nearest-neighbor interactions
but the constituent spins were restricted to $1$ and $\frac{1}{2}$.
Although an alignment of alternating spins $S$ and $s$ ($S>s$) in a
field, as described by the Hamiltonian
\begin{eqnarray}
   {\cal H}
    ={\displaystyle\sum_{j=1}^N}
    &\bigl[&
     (1+\delta)
     (\mbox{\boldmath$S$}_{j}\cdot\mbox{\boldmath$s$}_{j})_\alpha
    +(1-\delta)
     (\mbox{\boldmath$s$}_{j}\cdot\mbox{\boldmath$S$}_{j+1})_\alpha
    \nonumber \\
    &-&H(S_j^z+s_j^z)
    \bigr]\,,
   \label{E:H}
\end{eqnarray}
with
$(\mbox{\boldmath$S$}\cdot\mbox{\boldmath$s$})_\alpha
 =S^xs^x+S^ys^y+\alpha S^zs^z$,
is so interesting as to possibly exhibit a series of quantized
magnetization plateaux at
$m=\frac{1}{2} (1), \frac{3}{2} (2), \cdots, S+s-1$,
its magnetization curves have not systematically been studied so far.
In spite of the vigorous argument, there are few reports on
multi-plateau magnetization curves.
It is true that a double-plateau structure lies in NH$_4$CuCl$_3$
\cite{Kole81,Shir48}, but it is owing to the variety of exchange
interactions.
We here demonstrate that the ferrimagnetic chain (\ref{E:H})
generally exhibits {\it a $2s$-plateau magnetization curve without
any anisotropy and any bond polymerization}, namely, even at
$\alpha=1$ and $\delta=0$.
We believe that the present calculations will accelerate physical
measurements on vast ferrimagnetic chain compounds lying unexploited
in the field of both inorganic and organic chemistry.

   The ground state of the isotropic Hamiltonian (\ref{E:H}) without
the Zeeman term, which is a multiplet of spin $(S-s)N$, exhibits
elementary excitations of two distinct types \cite{Yama10}.
The excitations of ferromagnetic aspect, reducing the ground-state
magnetization, form a gapless dispersion relation, whereas those
of antiferromagnetic aspect, enhancing the ground-state
magnetization, are gapped from the ground state.
Therefore we can readily understand the initial step at $m=S-s$ in
the magnetization curve.
In the Ising limit $\alpha\rightarrow\infty$, the initial plateau is
nothing but the gapped excitation from the N\'eel-ordered state.
The classical gap-generation mechanism is unique.
Thus any magnetization curve in the Ising limit only has a single
plateau.
The scenario is qualitatively unchanged for an arbitrary $\alpha$
as long as we consider the classical vectors $\mbox{\boldmath$S$}_j$
and $\mbox{\boldmath$s$}_j$ of magnitude $S$ and $s$ instead of
quantum spins.
Therefore, the second and higher plateaux, if any, should generally
be based on a quantum mechanism.
\vskip 0mm
\begin{figure}
\begin{flushleft}
\quad\mbox{\psfig{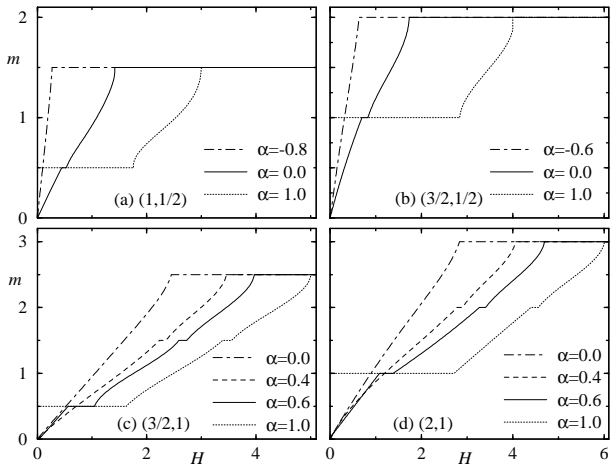}}
\end{flushleft}
\vskip 0mm
\caption{The ground-state magnetization curves for the Hamiltonian
         (2) with $\delta=0$ at various values of $\alpha$ and
         $(S,s)$.}
\label{F:m}
\end{figure}
\vskip -10mm
\begin{figure}
\begin{flushleft}
\qquad\mbox{\psfig{figure=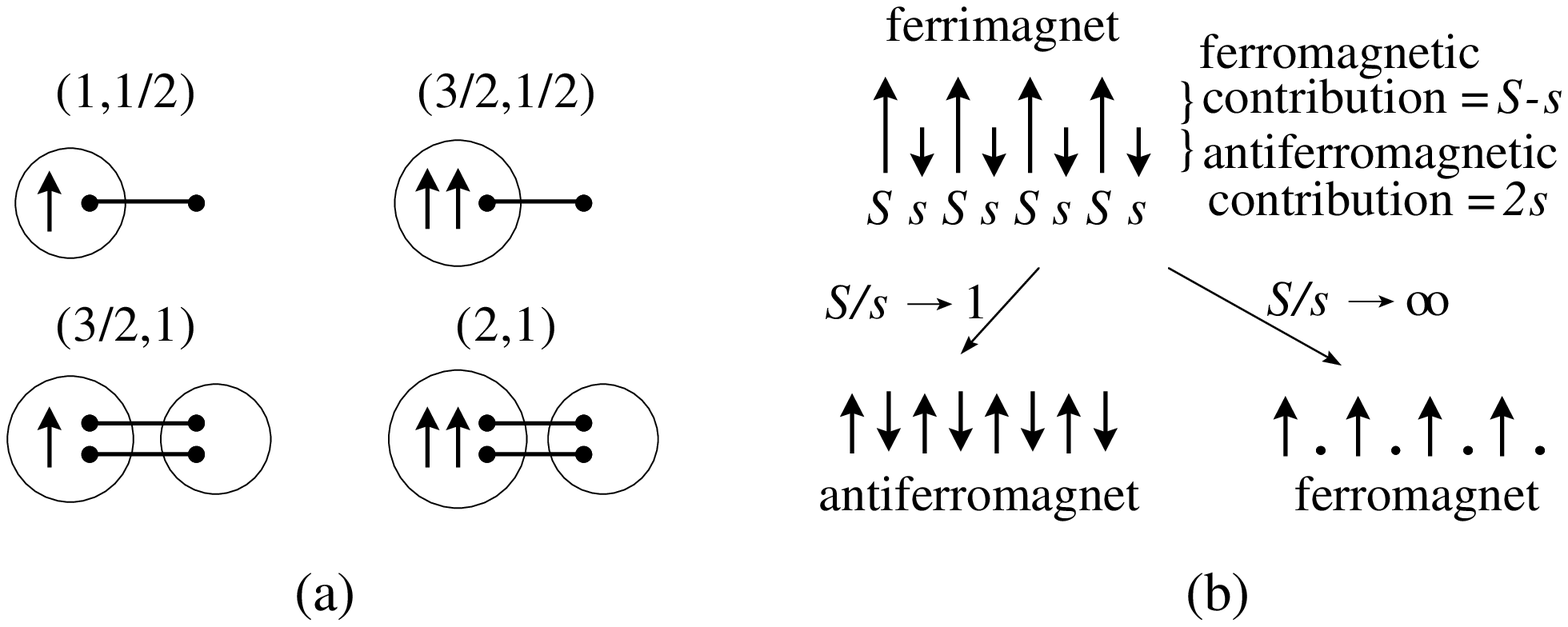,width=108mm,angle=0}}
\end{flushleft}
\vskip 0mm
\caption{(a) Schematic representations of the $M=(S-s)N$ ground
         states of spin-$(S,s)$ ferrimagnetic chains of $N$
         elementary cells in the decoupled-dimer limit.
         The arrow (the bullet symbol) denotes a spin $\frac{1}{2}$
         with its fixed (unfixed) projection value.
         The solid segment is a singlet pair.
         (b) A ferrimagnet may be regarded as the combination of a
         ferromagnet and an antiferromagnet, where $S-s$ and $2s$ of
         the total spin amplitude $S+s$ play the ferromagnetic and
         antiferromagnetic roles, respectively.}
\label{F:illust}
\end{figure}

\section{Numerical Procedure}

   We perform a scaling analysis \cite{Saka53} on the numerically
calculated energy spectra of finite clusters up to $N=12$.
With $E(N,M)$ being the lowest energy in the subspace with a fixed
magnetization $M$ for the Hamiltonian (\ref{E:H}) without the Zeeman
term, the upper and lower bounds of the field which induces the
ground-state magnetization $M$ are expressed as
\begin{equation}
   H_\pm(N,M)=\pm E(N,M\pm 1)\mp E(N,M)\,.
\end{equation}
The length of the plateau with the unit-cell magnetization
$m\equiv M/N$ is obtained as
\begin{equation}
   {\mit\Delta}_N(m)=H_+(N,M)-H_-(N,M)\,.
\end{equation}
Therefore, calculating $E(N,M)$ at each sector of $M$ and
extrapolating the resultant $H_\pm(N,M)$ with respect to $N$, we can
obtain the thermodynamic-limit magnetization curves.
Since the correlation length of the present system is considerably
small \cite{Pati94,Breh21}, this scaling analysis works very well.
The precision of the obtained magnetization curves is generally down
to three decimal places.
There is at most slight uncertainty in the second decimal place.

\section{Results and Discussion}

   We show in Fig. \ref{F:m} the thus-obtained magnetization curves.
Making use of the Schwinger boson representation:
\begin{equation}
   \left.
   \begin{array}{ll}
      S_j^+=A_j^\dagger B_j\,,&
      S_j^z=\frac{1}{2}(A_j^\dagger A_j-B_j^\dagger B_j)\,,\\
      s_j^+=a_j^\dagger b_j\,,&
      s_j^z=\frac{1}{2}(a_j^\dagger a_j-b_j^\dagger b_j)\,,
   \end{array}
   \right.
   \label{E:SchBoson}
\end{equation}
the $M=S-s$ ground state of the decoupled dimers ($\delta=1$) are
described as
$\prod_j
 (A_j^\dagger)^{S-s}
 (A_j^\dagger b_j^\dagger-B_j^\dagger a_j^\dagger)^{2s}
 |0\rangle$,
whose schematic representation is given in Fig. \ref{F:illust}(a).
Therefore, any plateau is enhanced by the bond alternation and the
magnetization curve ends up with $2s$ steps, which are attributable
to the crackion excitations \cite{Knab27,Fath83}, in the
decoupled-dimer limit.
Hence we here concentrate on the uniform-bond case ($\delta=0$).
Surprisingly, in a certain region of $\alpha$ including the
Heisenberg point, the spin-$(S,s)$ chain generally possesses a
$2s$-plateau magnetization curve.
To the best of our knowledge, this is the first report on the
multi-plateau structure depending on neither anisotropy nor bond
polymerization.
$2s$ plateaux appear, but still, that does not mean the plateaux are
dominated only by the smaller spin.
Ferrimagnets have both ferromagnetic and antiferromagnetic features
\cite{Yama08}.
The mixed aspect is explicitly exhibited, for instance, in their
thermodynamics, where the specific heat and the magnetic
susceptibility times temperature behave like $T^{1/2}$ and $T^{-1}$
at low temperatures, respectively, whereas they exhibit a
Schottky-like peak and a round minimum at intermediate temperatures.
Figure \ref{F:illust}(b), as well as Fig. \ref{F:illust}(a), shows
that the spin amplitude $S-s$ plays the ferromagnetic role, while
$2s$ plays the antiferromagnetic one \cite{Yama24}.
Considering that any magnetization plateau originates from an
antiferromagnetic interaction, it is convincing that $2s$ of the
total spin amplitude $S+s$ contributes to the plateaux appearing.

   Let us turn back to Fig. \ref{F:m} and observe the plateaux more
carefully, especially as functions of $\alpha$.
In the cases of $s=\frac{1}{2}$, the plateaux are quite tough against
the $XY$-like anisotropy.
They are stable all over the antiferromagnetic-coupling region.
These observations are in contrast with the classical behavior.
The classical spin-$(S,s)$ Heisenberg Hamiltonian also exhibits the
magnetization plateau at $m=S-s$, but it survives only a small amount
of $XY$-like anisotropy.
For instance, the critical value for $(S,s)=(1,\frac{1}{2})$ is
estimated as $\alpha_{\rm c}=0.943(1)$ \cite{Saka53}.
The contrast between quantum spins and classical vectors suggests
that the single plateaux in the quantum-spin magnetization curves may
be attributed to the valence-bond excitation gap
({\it valence-bond gap}) rather than the N\'eel-state excitation gap
({\it N\'eel gap}).
From this point of view, the $\alpha$-dependences of the two
coexistent plateaux in the cases of $s=1$ are interesting.
The tiny second plateaux are much more stable against the $XY$-like
anisotropy than the steady-looking initial steps.
Since the N\'eel state reaches the saturation via a single-step
excitation, the second and higher plateaux should originate from the
valence-bond gap.
The magnitude of the gap exponentially decreases with the increase of
$m$, but the quantum gap-generation mechanism itself is rather tough
against the $XY$-like anisotropy.
The initial plateaux in the multi-step magnetization curves, which
are relatively unstable against the $XY$-like anisotropy, may be
attributed to the N\'eel gap.
It seems that the lowest-lying magnetization plateaux are of quantum
appearance for $s=\frac{1}{2}$ but are of classical aspect for
$s\geq 1$.
\vskip 2mm
\begin{figure}
\begin{flushleft}
\qquad\mbox{\psfig{figure=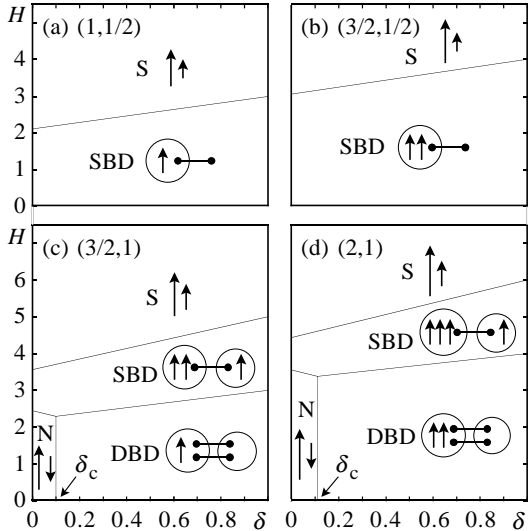,width=70mm,angle=0}}
\end{flushleft}
\vskip 0mm
\caption{The variational ground-state phase diagrams on the
         $\delta H$-plane for the Hamiltonian (2) with $\alpha=1$ at
         various values of $(S,s)$.}
\label{F:VPhD}
\end{figure}

   In order to characterize the plateaux, we introduce a variational
wave function for the ground state of the model (\ref{E:H}) as
\begin{eqnarray}
   &&|{\rm g}\rangle
   =c_{\rm N}\prod_{j=1}^N
    (A_j^\dagger)^{2S}(b_j^\dagger)^{2s}|0\rangle
   \nonumber \\
   &&\ +
    \sum_{l=0}^{2s}c_{\rm VB}^{(l)}
    \prod_{j=1}^N
    (A_j^\dagger)^{2S-l}(a_j^\dagger)^{2s-l}
    (A_j^\dagger b_j^\dagger-B_j^\dagger a_j^\dagger)^l|0\rangle\,,
   \label{E:VWF}
\end{eqnarray}
where $c_{\rm N}$ and $c_{\rm VB}^{(l)}$ are the mixing coefficients.
Since all the elemental states are asymptotically orthogonal to each
other, the thermodynamic-limit variational ground states are
considerably simple, as shown in Fig. \ref{F:VPhD}, where we consider
the Heisenberg point.
The phase diagrams are exact on the line of $\delta=1$, where the
spin-$(\frac{3}{2},1)$ chain, for example, reaches the saturation (S)
via the double-bond dimer (DBD) and single-bond dimer (SBD) states.
The point is that the $\delta=0$ ground states are better
approximated by the decoupled-dimer states than by the N\'eel (N)
states in the cases of $s=\frac{1}{2}$, while vice versa in all other
cases.
For $s=\frac{1}{2}$, the variational wave function (\ref{E:VWF}) ends
up with $c_{\rm N}=0$ all over the $\delta$-$H$ plain.
The N\'eel-dimer crossover point $\delta_{\rm c}$ is given by
\begin{equation}
   \delta_{\rm c}=\frac{2Ss-A+B+C}{A-B+C}\,,
\end{equation}
with
\begin{eqnarray}
   &&
   A=\sum_{l=1}^{2s}
     \biggl\{
      \frac{(2S-2s+l)!(2S-2s+l-1)!}{[(2S-2s)!]^2l!(l-1)!}
   \nonumber \\
   &&\qquad
      \times[S(S+1)-(S-2s+l-1)(S-2s+l)]
   \nonumber \\
   &&\qquad
      \times[s(s+1)-(s-l)(s-l+1)]
     \biggr\}^{1/2}
   \nonumber \\
   &&\qquad
    /\sum_{l=0}^{2s}
      \frac{(2S-2s+l)!}{(2S-2s)!l!}\,,
   \nonumber \\
   &&
   B=\sum_{l=0}^{2s}
      \frac{(2S-2s+l)!}{(2S-2s)!l!}(S-2s+l)(s-l)
   \nonumber \\
   &&\qquad
    /\sum_{l=0}^{2s}
      \frac{(2S-2s+l)!}{(2S-2s)!l!}\,,
   \nonumber \\
   &&
   C=\sum_{l=0}^{2s}
      \frac{(2S-2s+l)!}{(2S-2s)!l!}(S-2s+l)
   \nonumber \\
   &&\qquad
      \times
     \sum_{l=0}^{2s}
      \frac{(2S-2s+l)!}{(2S-2s)!l!}(s-l)
   \nonumber \\
   &&\qquad
    /\sum_{l=0}^{2s}
     \frac{(2S-2s+l)!}{(2S-2s)!l!}\,.
\end{eqnarray}
\begin{figure}
\begin{flushleft}
\qquad\quad\ \mbox{\psfig{figure=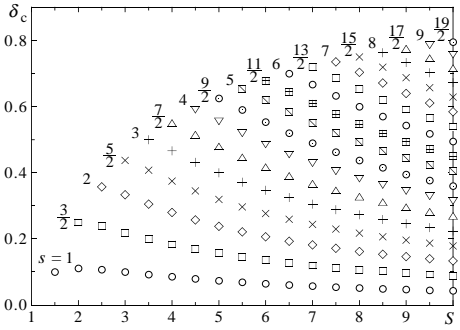,width=74mm,angle=0}}
\end{flushleft}
\vskip 0mm
\caption{The variational crossover points $\delta_{\rm c}$ between
         the N\'eel-ordered state and the $2s$-bond decoupled-dimer
         state for the Hamiltonian (2) with $\alpha=1$ at various
         values of $(S,s)$.}
\label{F:deltac}
\end{figure}
\noindent
$\delta_{\rm c}$ does not exist for
$s=\frac{1}{2}$, whereas it is an increasing function of $s$ for
$s\geq 1$, as shown in Fig. \ref{F:deltac}.
Thus, the N\'eel-gap-like character of the initial plateaux in the
multi-step magnetization curves become more and more settled as $s$
increases, which leads to the instability of the plateaux against the
$XY$-like anisotropy.
The exceptional cases of $s=\frac{1}{2}$ may be recognized as the
quantum limit.
Another simple calculation also supports this scenario.
Let us consider a spin-wave description and a perturbation
treatment of the antiferromagnetic excitation gap from the ground
state.
The spin-wave excitations are based on the N\'eel-order background,
whereas the perturbation from the decoupled-dimer limit assumes the
crackion-like excitations to appear in the valence-bond background.
We compare in Table \ref{T:Delta} both estimates with the exact
values, namely, the upper critical fields for the initial plateaux.
In the cases of $s=\frac{1}{2}$, the perturbation calculations  are
better than the spin-wave estimates, while vice versa in the cases
of $s\geq 1$.
We are again convinced that the single plateaux for $s=\frac{1}{2}$
are relatively of quantum aspect, while the lowest-magnetization
plateaux in the multi-step process for $s\geq 1$ are relatively of
classical aspect.
All other plateaux, the second and higher steps, should essentially
be based on the quantum mechanism.
Now here is a question:
As $\alpha$ increases, at which point do the quantum plateaux for
$m>S-s$ disappear?
Our numerical investigations estimate that they survive the whole
region of $\alpha>1$ and disappear in the Ising limit.

\begin{figure}
\begin{flushleft}
\qquad\quad\ \mbox{\psfig{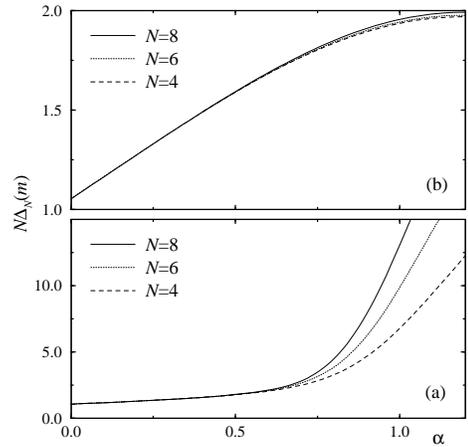}}
\end{flushleft}
\vskip 0mm
\caption{Scaled quantity $N{\mit\Delta}_N(m)$ versus $\alpha$ at
         $m=\frac{1}{2}$ (a) and $m=\frac{3}{2}$ (b) in the case of
         $(S,s)=(\frac{3}{2},1)$.}
\label{F:NDelta}
\end{figure}

   Finally we draw the ground-state phase diagrams on the
$\alpha\delta$-plane.
If the system is massive at the sector of magnetization $M$,
$H_\pm(N,M)$ are extrapolated to different thermodynamic-limit values
$H_\pm(m)$ with exponential size corrections.
On the other hand, in the critical phase, $H_\pm(N,M)$ converge to
the same value as \cite{Card85,Affl46}
\begin{equation}
   H_{\pm}(N,M)
    \sim H(m) \pm \frac{\pi v_{\rm s}\eta}{N}\,,
    \label{E:CFT}
\end{equation}
where $v_{\rm s}$ is the spin-wave velocity and $\eta$ is the
critical index for the relevant spin-correlation function.
Therefore, we can visualize the phase transition by plotting the
scaled gap $N{\mit\Delta}_N(m)$ as a function of $\alpha$ as shown
in Fig. \ref{F:NDelta}.
The phase boundaries could in principle be extracted
\vskip 0mm
\vskip 0mm
\begin{figure}
\begin{flushleft}
\quad\mbox{\psfig{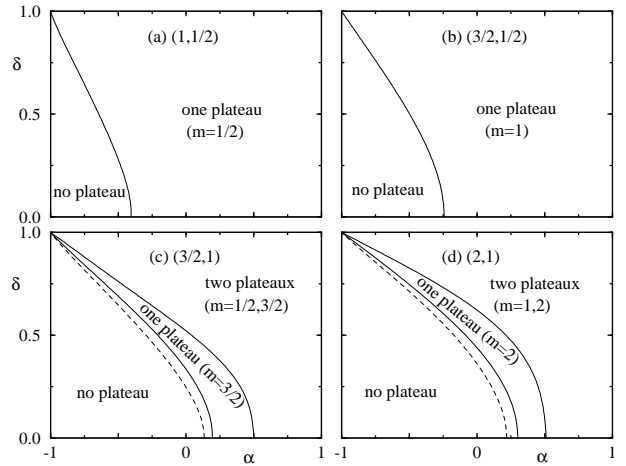}}
\end{flushleft}
\vskip 0mm
\caption{The ground-state phase diagrams on the $\alpha\delta$-plane
         for the Hamiltonian (2) at various values of $(S,s)$.}
\label{F:PhD}
\end{figure}
\vskip 2mm
\noindent
from the phenomenological renormalization-group equation
\cite{Nigh61} taking ${\mit\Delta}_N(m)$ as the order parameter.
However, the anisotropy-induced breakdown of the plateau is a
transition of the Kosterlitz-Thouless type and the fixed point
could only be determined with great uncertainty \cite{Nomu23}.
Thus we here rely upon the critical exponent $\eta$ which should
cross over the value $\frac{1}{4}$ on the phase boundary.
Provided $v_{\rm s}$ is given, we can estimate $\eta$ using the
scaling law (\ref{E:CFT}).
We obtain $v_{\rm s}$ directly from the dispersion relation.
Using the scaling relation \cite{Card85,Affl46}
\begin{equation}
   \frac{E(N,M)}{N}
    \sim \varepsilon(m)-\frac{\pi cv_{\rm s}}{N^2}\,,
\end{equation}
we further verify the central charge being unity in the critical
region, though it is not so useful in determining the phase boundary.
The thus-obtained phase boundaries are shown by solid lines in Fig.
\ref{F:PhD}.
The single plateaux, with a quantum base, are stable over the whole
antiferromagnetic-coupling region (a,b), while the initial plateaux
in the multi-step process, taking on a classical character, less
survive the $XY$-like anisotropy than the quantum higher plateaux
(c,d).
The existence of the second plateaux without any bond polymerization,
which is the main issue in the present article, should be verified
very carefully.
So then we further employ an idea of the level spectroscopy
\cite{Okam33}, thus called, in analyzing the second plateau.
Comparing the relevant excitation energies whose scaling dimensions
are $2$ at the critical point, we recognize the level crossing of
them as the phase boundary.
The thus-detected transitions are also plotted by broken lines in
Fig. \ref{F:PhD}.
The slight difference between the two estimates inevitably arises
from the logarithmic corrections to the scaling law (\ref{E:CFT}),
where the level spectroscopy is more reliable than the naivest
scaling analysis.
Anyway, we may now fully be convinced of the existence of the novel
multi-plateau magnetization curves.

\section{Summary and Future Aspect}

   The one-dimensional Heisenberg ferrimagnet with alternating spins
$S$ and $s$ exhibits a $2s$-plateau magnetization curve even at the
most symmetric point.
It is interesting to compare the present spin-$(S,s)$ ferrimagnetic
chain with the spin-$\frac{1}{2}$ bond-polymerized chain of period
$2(S+s)$, that is, the $
(2S-1)$-times-ferromagnetic-antiferromagnetic-$
(2s-1)$-times-ferromagnetic-antiferromagnetic chain.
In the strong ferromagnetic-coupling limit, the latter may be
regarded as equivalent to the former.
In the same meaning, the spin-$S$ antiferromagnetic Heisenberg chain
can be viewed as a spin-$\frac{1}{2}$ bond-polymerized chain of
period $2S$.
Such replica chains generally exhibit magnetization plateaux in
certain regions of the ratio of the ferromagnetic coupling
$J_{\rm F}$ to the antiferromagnetic one $J_{\rm A}$,
$\gamma\equiv J_{\rm F}/J_{\rm A}$.
However, in the case of the spin-$\frac{1}{2}$
ferromagnetic-ferromagnetic-antiferromagnetic chain which is the
replica model of the spin-$\frac{3}{2}$ antiferromagnetic Heisenberg
chain, Okamoto \cite{Okam39} and Hida \cite{Hida59} reported that the
plateau vanishes at $\gamma=4\sim 5$ and therefore the pure
Heisenberg chain exhibits no plateau.
Thus the plateaux observed here in the most symmetric Heisenberg
ferrimagnets still interest us to a great extent.

   As we come up the steps, the plateau length exponentially
decreases.
It is hard to numerically observe the higher-lying plateaux, still
harder experimentally.
Only the first and second plateaux may lie within the limits of
measurement.
In this context, we are fortunate to have a series of bimetallic
quasi-one-dimensional complexes MM$'$(EDTA)$\cdot$6H$_2$O
($\mbox{M},\mbox{M}'=\mbox{Mn},\mbox{Co},\mbox{Ni},\mbox{Cu}$)
\cite{Dril53}.
Their exchange coupling constants are all about $10k_{\rm B}$[K]
and thus the complete magnetization curves could technically be
observed.
Magnetization measurements on them, especially with
$\mbox{M}=\mbox{Mn}\,(S=\frac{5}{2}), \mbox{Co}\,(S=\frac{3}{2})$ and
$\mbox{M}'=\mbox{Ni}\,(s=1)$, are encouraged.
The plateaux would more or less be obscured in any actual
measurement, but they, however small, should necessarily be detected
by some anomaly in the magnetic susceptibility.
The chemical modification of the bond alternation $\delta$ and/or the
exchange anisotropy $\alpha$ must help us to directly observe the
second-step plateaux, though we take main interest in the Heisenberg
point.

\acknowledgments

   The authors thank Dr. K. Okamoto for useful discussion.
This work is supported by the Japanese Ministry of Education,
Science, and Culture through Grant-in-Aid No. 11740206 and by the
Sanyo-Broadcasting Foundation for Science and Culture.
The computation was done in part using the facility of the
Supercomputer Center, Institute for Solid State Physics, University
of Tokyo.

\begin{table}
\caption{Estimates of the antiferromagnetic excitation gap from the
         ground state for the Hamiltonian (2) with $\alpha=1$ and
         $\delta=0$ by the use of the linear spin-wave theory
         (spin wave), the perturbation from the decoupled-dimer limit
         (perturbation), and the exact diagonalization (exact).}
\begin{tabular}{cccl}
 $(S,s)$ & spin wave & perturbation & exact \\
\tableline
\noalign{\vskip 1mm}
$(          1,\frac{1}{2})$ & $1$ & $\frac{11}{ 9}$ & $1.759( 1)$ \\
\noalign{\vskip 1mm}
$(\frac{3}{2},\frac{1}{2})$ & $2$ & $\frac{17}{ 8}$ & $2.842( 1)$ \\
\noalign{\vskip 1mm}
$(\frac{3}{2},          1)$ & $1$ & $\frac{16}{45}$ & $1.615(10)$ \\
\noalign{\vskip 1mm}
$(          2,          1)$ & $2$ & $\frac{11}{30}$ & $2.730( 5)$ \\
\end{tabular}
\label{T:Delta}
\end{table}
\widetext
\end{document}